\documentclass[
prc,%
10pt,%
final,%
notitlepage,%
oneside,%
twocolumn,%
nobibnotes,%
nofootinbib,% In the current version of REVTeX, when this option is on,
superscriptaddress,%
floatfix,%
floatfix,%
showkeys,%
showpacs]%
{revtex4}
\usepackage{color}%\textcolor{red}{}
\usepackage{amsfonts}
\usepackage{amsbsy}
\usepackage{mathrsfs}
\usepackage{graphicx}
\def\lsim{\mathrel{\rlap{
\lower4pt\hbox{\hskip-3pt$\sim$}}
    \raise1pt\hbox{$<$}}}     %less than approx. symbol
\def\gsim{\mathrel{\rlap{
\lower4pt\hbox{\hskip-3pt$\sim$}}
    \raise1pt\hbox{$>$}}}     %greater than or approx. symbol

\begin{document}
\title{
Vorticity in heavy-ion collisions at the JINR Nuclotron-based Ion Collider fAcility} 
\author{Yu. B. Ivanov}\thanks{e-mail: Y.Ivanov@gsi.de}
\affiliation{National Research Centre "Kurchatov Institute", 123182 Moscow, Russia} 
\affiliation{National Research Nuclear University "MEPhI"  (Moscow Engineering
Physics Institute),
%Kashirskoe sh. 31, 
Moscow 115409, Russia}
\author{A. A. Soldatov}\thanks{e-mail: saa@ru.net}
\affiliation{National Research Nuclear University "MEPhI"  (Moscow Engineering
Physics Institute),
%Kashirskoe sh. 31, 
Moscow 115409, Russia}
\begin{abstract}
Vorticity of matter generated in noncentral heavy-ion collisions at energies of 
the  Nuclotron-based Ion Collider fAcility (NICA) at the 
Joint Institute for Nuclear Research (JINR) in Dubna is studied. 
Simulations are performed within the model of the three-fluid dynamics
(3FD) which reproduces the major part of bulk observables at these energies. 
Comparison with earlier calculations is done. Qualitative pattern of the 
vorticity evolution is analyzed. It is demonstrated that the vorticity is 
mainly located at the border between participants and spectators. 
In particular, this implies that 
the relative $\Lambda$-hyperon polarization should be stronger at rapidities
of the fragmentation regions  
than that in the midrapidity region. 
\end{abstract}
\maketitle
%\today

\section{Introduction}

In peripheral  collisions of high-energy heavy ions the system
has a large angular momentum \cite{Becattini:2007sr} that 
may result in observable consequences.  
The large angular momentum
can manifest itself in 
a chiral vortical effect that
results in induced currents and charge separation
\cite{Kharzeev:2007tn,Rogachevsky:2010ys} similarly to the so-called chiral magnetic
effect \cite{Kharzeev:2004ey,Fukushima:2008xe,Zhitnitsky:2013owa}. 
Another possible manifestation is 
the polarization of secondary
produced particles \cite{Becattini:2007sr,Liang:2004ph,Betz:2007kg,Karpenko:2016jyx,Xie:2016fjj}. 
Preliminary experimental results on hyperon polarization in 
heavy-ion collisions at energies of the Beam Energy
Scan (BES) program at Relativistic Heavy Ion Collider (RHIC)
at Brookhaven %National Laboratory (BNL) 
has been recently reported \cite{Upsal:2016phr}.
Experimental observation of these effects may give us 
additional information on the dynamics the  heavy-ion collisions,   
e.g. on possible  Kelvin-Helmholtz instability 
\cite{Csernai:2011qq,Wang:2013xtp} or other turbulent
phenomena \cite{Florchinger:2011qf}.

The vorticity developed in heavy-ion collisions was estimated within various models. 
These estimates mainly concern the energy of Large Hadron Collider (LHC)  
and RHIC energies %of BNL Relativistic Heavy-Ion Collider (RHIC) 
\cite{Huang:2011ru,Csernai:2013bqa,Becattini:2015ska,Jiang:2016woz,Deng:2016gyh}. 
A comprehensive study of $\Lambda$ polarization at 
BES-RHIC energies have recently been done in Ref. 
\cite{Karpenko:2016jyx} which only partially overlap with the NICA energy range. 
Only two recent studies \cite{Csernai:2014ywa,Teryaev:2015gxa}
are dedicated to lower energies of NICA at JINR. 
However, those studies \cite{Csernai:2014ywa,Teryaev:2015gxa} 
were performed within different approaches 
(\cite{Csernai:2014ywa} within the relativistic PICR hydro approach \cite{Csernai:2013bqa}
and \cite{Teryaev:2015gxa} within the hadron-string dynamics model \cite{Cassing99}) 
and for nuclear collisions at different collision energies which makes 
difficult their direct comparison. Very recently the approach of Ref. \cite{Csernai:2013bqa} 
was further developed to estimate the $\Lambda$ polarization at the top NICA energies 
\cite{Xie:2016fjj}.

In the present paper the vorticity is simulated within the 3FD model  
\cite{3FD} 
for several collision energies in the NICA energy range. 
{
This study is also relevant to the recently announced 
STAR Fixed-Target Program at RHIC \cite{Meehan:2016qon}.  
}
The 3FD model is quite successful in reproduction 
of the major part of bulk
observables: the baryon stopping \cite{Ivanov:2013wha,Ivanov:2012bh}, 
yields of different hadrons, their rapidity and transverse momentum
distributions \cite{Ivanov:2013yqa,Ivanov:2013yla}, 
the elliptic \cite{Ivanov:2014zqa} 
and directed 
\cite{Konchakovski:2014gda} flow excitation functions. 
Therefore, it would be instructive to compare the 3FD vorticity pattern 
with those in above mentioned approaches \cite{Csernai:2014ywa,Teryaev:2015gxa}.

\section{Vorticity in the 3FD model}
\label{Model}

There are several definitions of the vorticity used in the literature that are suitable 
for analyzing different aspects of the rotation effects. In the present study we  
consider two of them. The first one is the relativistic kinematic vorticity
   \begin{eqnarray}
   \label{rel.kin.vort.}
   \omega_{\mu\nu} = \frac{1}{2}
   (\partial_{\nu} u_{\mu} - \partial_{\mu} u_{\nu}), 
   \end{eqnarray}
where $u_{\mu}$ is a collective local four-velocity of the matter. 
This type of the vorticity is directly relevant to the chiral vortical effect 
\cite{Rogachevsky:2010ys}
that is caused by coupling to medium vorticity and leads to contribution
to the electromagnetic current
   \begin{eqnarray}
   \label{Jem}
J^\kappa_{\rm{e}} = \frac{N_c}{4\pi^2 N_f} \varepsilon^{\kappa\lambda\mu\nu}
\partial_{\mu} u_{\nu} \; \partial_\lambda
\left(\theta \sum_j e_j \mu_j \right), 
   \end{eqnarray}
where $N_c$ and $N_f$ are the number of colors and flavors
respectively, $e_j$ and $\mu_j$ are the electric charge and chemical
potential of a particle of $j$ flavor, respectively,  
and $\theta$ is the topological QCD field.

Another type of the vorticity is so-called thermal vorticity 
   \begin{eqnarray}
   \label{therm.vort.}
%   \bar{\omega}_{\mu\nu} = \frac{1}{2}
   \varpi_{\mu\nu} = \frac{1}{2}
   (\partial_{\nu} \hat{\beta}_{\mu} - \partial_{\mu} \hat{\beta}_{\nu}), 
   \end{eqnarray}
where $\hat{\beta}_{\mu}=\hbar\beta_{\mu}$ and $\beta_{\mu}=u_{\nu}/T$ 
with $T$ being the local temperature. Thus, $\varpi$ is dimensionless. 
It is directly related to the polarization vector, $\Pi^\mu(p)$, of a spin 1/2 particle
in a relativistic fluid \cite{Becattini:2013fla} 
   \begin{eqnarray}
%   \begin{equation}
\label{polint}
 \Pi^\mu(p)=\frac{1}{8m} \frac{\int_\Sigma \mathrm{d} \Sigma_\lambda p^\lambda
   n_F (1-n_F) \: p_\sigma \epsilon^{\mu\nu\rho\sigma} \partial_\nu \beta_\rho}
 {\int_\Sigma \Sigma_\lambda p^\lambda \, n_F},
%   \end{equation}
   \end{eqnarray}
where $n_F$ is the Fermi-Dirac-Juttner distribution function and the 
integration runs over the freeze-out hypersurface $\Sigma$.

Unlike the conventional hydrodynamics, where local
instantaneous stopping of projectile and target matter is
assumed, a specific feature of the 3FD description \cite{3FD} 
is a finite stopping power resulting in a counterstreaming
regime of leading baryon-rich matter. This generally
nonequilibrium regime of the baryon-rich matter
is modeled by two interpenetrating baryon-rich fluids 
initially associated with constituent nucleons of the projectile
(p) and target (t) nuclei. In addition, newly produced particles,
populating the midrapidity region, are associated with a fireball
(f) fluid.
At later stages the baryon-rich and fireball fluids may consist
of any type of hadrons and/or partons (quarks and gluons),
rather than only nucleons and pions.
Each of these fluids is governed by conventional hydrodynamic equations 
coupled by friction terms in the right-hand sides of the Euler equations. 
These friction terms describe energy--momentum loss of the 
baryon-rich fluids. 
%A part of this
%loss is transformed into thermal excitation of these fluids, while another part 
%gives rise to particle production into the fireball fluid.

Thus, the system is characterized by three hydrodynamical velocities,  
$u_{\alpha}^{\mu}$ with $\alpha=$ p, t and f, attributed to these fluids. 
At NICA energies the interpenetration of the p and t  fluids takes place 
only at the initial stage of the nuclear collision. At later stages 
a complete mutual stopping occurs and these fluids get unified.  Therefore, 
we define a collective 4-velocity of the baryon-rich matter associating it 
with the total baryon current 
   \begin{eqnarray}
   \label{bar-u}
   u^{\mu}_B =  J_{B}^{\mu}/|J_{B}|, 
   \end{eqnarray}
where $J_{B}^{\mu} = n_{p}u_{p}^{\mu}+n_{t}u_{t}^{\mu}$ is the baryon
current defined in terms of proper baryon densities $n_{\alpha}$ and
hydrodynamic 4-velocities $u_{\alpha}^{\mu}$, and 
{
   \begin{eqnarray}
   \label{nb-prop}
   |J_{B}|= \left(J_{B}^{\mu} J_{B\mu}\right)^{1/2}\equiv n_B
   \end{eqnarray}
is the proper (i.e. in the local rest frame) baryon density of the p and t  fluids. 
In particular, this proper baryon density allows us to construct a simple fluid 
unification measure 
   \begin{eqnarray}
   \label{unification}
   1-\frac{n_{p}+n_{p}}{n_B}
   \end{eqnarray}
which is zero, when the p and t  fluids are mutually stopped and unified, and  
has a positive value increasing  with rise of
the relative velocity of the p and t  fluids. 
}

The energy accumulated by the fireball fluid is an order of magnitude lower than that 
in the baryon-rich fluids even at $\sqrt{s_{NN}}=$ 9.2 GeV, i.e. the top 
NICA energy. Therefore, we concentrate on the vorticity of the baryon-rich fluids. 
Thus, the vorticities of Eqs. (\ref{rel.kin.vort.}) and (\ref{therm.vort.})
are considered in terms of $u^{\mu}_B$. 

The temperature, $T_B$, that is required in calculations of the 
thermal vorticity (\ref{therm.vort.}), 
also needs some comments. It is defined as a local  proper-energy-density-weighted temperature
   \begin{eqnarray}
   \label{Tm-dissipative}
T_B = \sum_{\alpha=p,t} T_{\alpha} \varepsilon_{\alpha} \Big/ \sum_{\alpha=p,t} \varepsilon_{\alpha} 
   \end{eqnarray}
where $\varepsilon_{\alpha}$ is the proper energy density of the $\alpha$ fluid. 
At the initial nonequilibrium stage of the collision  
[i.e. at $t\lsim 4$ fm/c for 4.9 GeV and $t\lsim 2$ fm/c for 7.7 GeV for midcentral Au+Au collisions 
considered below] 
this quantity does not relate 
to a true temperature of the system just because the temperature concept is 
inapplicable to a strongly nonequilibrium system. However, this temperature 
is close to the true temperature of the system at the expansion stage of the collision, 
when baryon-rich fluids are practically unified.

The physical input of the present 3FD calculations is described in
Ref.~\cite{Ivanov:2013wha}. The friction between fluids was fitted to reproduce
the stopping power observed in proton rapidity distributions for each EoS, 
as it is described in  Ref. \cite{Ivanov:2013wha} in detail.
The simulations in 
\cite{Ivanov:2013wha,Ivanov:2012bh,Ivanov:2013yqa,Ivanov:2013yla,Ivanov:2014zqa,Konchakovski:2014gda} 
were performed with different 
equations of state (EoS's)---a purely hadronic EoS \cite{gasEOS}  
and two versions of the EoS involving the   deconfinement
 transition \cite{Toneev06}, i.e. a first-order phase transition  
and a smooth crossover one. In the present paper we demonstrate results with 
only the crossover EoS as the most successful in reproduction of various 
bulk observables.  

\begin{figure}[htb]
%\vspace*{-14mm}
\includegraphics[width=8.2cm]{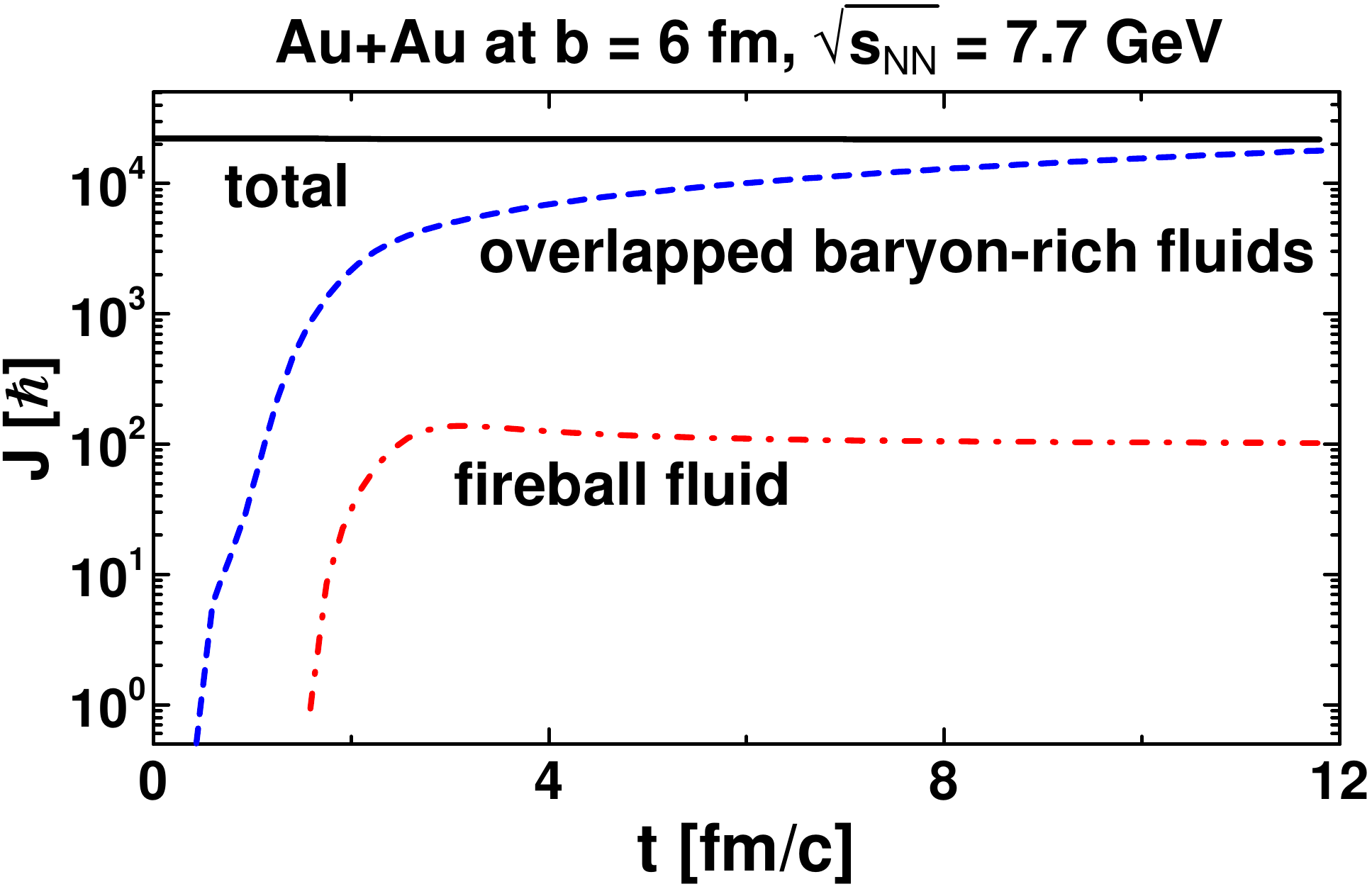}
 \caption{(Color online)
Time evolution of the total angular momentum (conserved quantity), 
the angular momentum of the baryon-rich fluids in their overlap 
region and the angular momentum of the fireball fluid 
in the semi-central ($b=$ 6 fm) Au+Au collision at $\sqrt{s_{NN}}=$ 7.7 GeV. 
Calculations are done with the crossover EoS. 
}
\label{fig0}
\end{figure}
\begin{figure*}[!htb]
%\vspace*{-14mm}
\includegraphics[width=18.cm]{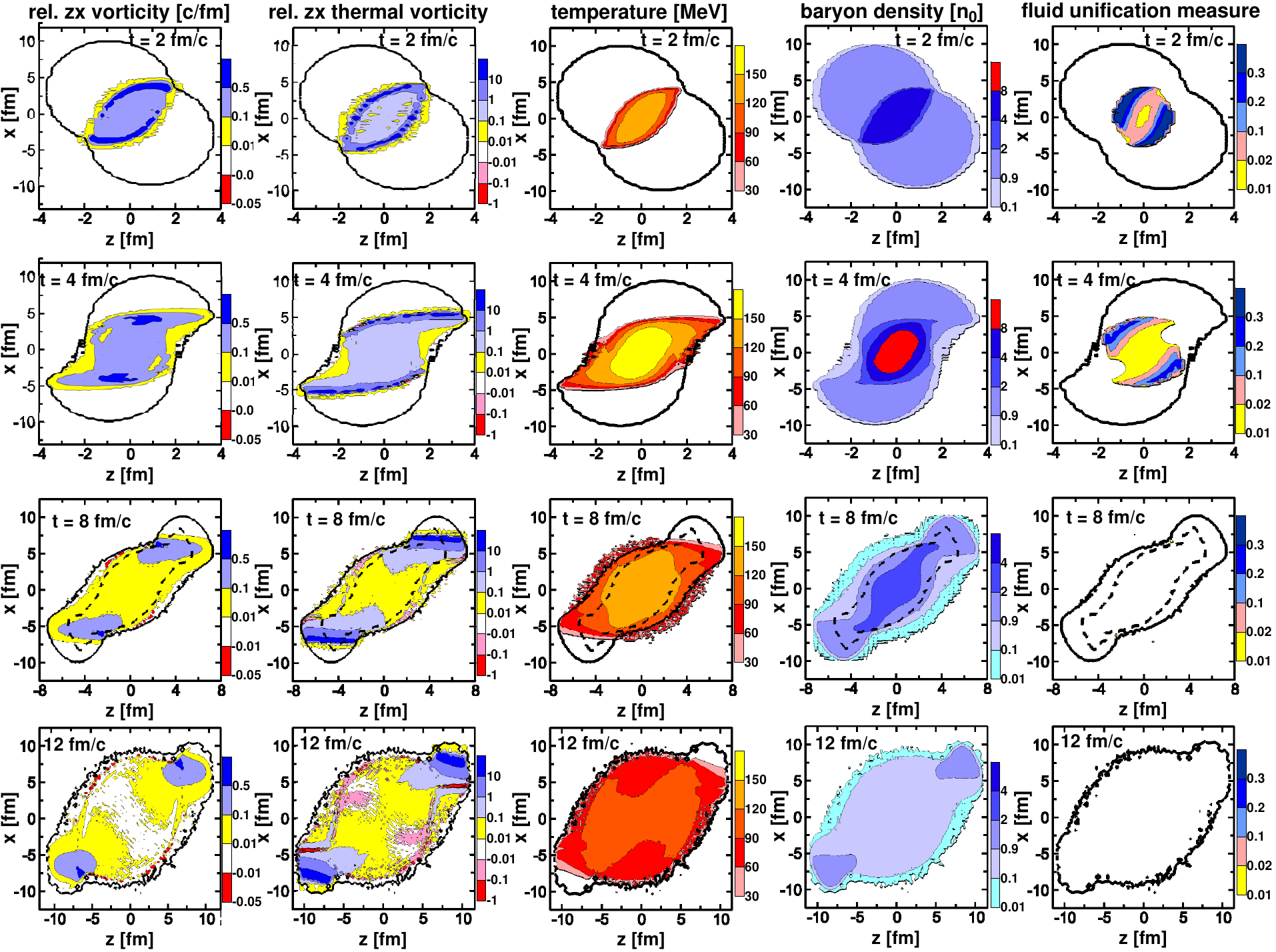}
 \caption{(Color online)
{
Columns from left to right: The proper-energy-density weighted 
relativistic kinematic $zx$ vorticity, the thermal $zx$ vorticity, 
the temperature [cf. Eq. (\ref{Tm-dissipative})], the proper baryon density ($n_B$)  
[cf. Eq. (\ref{nb-prop})] in units of the the normal nuclear density ($n_0=0.15$ 1/fm$^3$),
and the fluid unification measure [cf. Eq. (\ref{unification})]
}
of the baryon-rich subsystem,  
in the reaction plane at various time instants 
in the semi-central ($b=$ 6 fm) Au+Au collision at $\sqrt{s_{NN}}=$ 4.9 GeV. 
Calculations are done with the crossover EoS. $z$ axis is the 
beam direction. Note different scale along the $z$ axis at different time instants. 
The outer bold solid contour displays the border of the baryon-rich matter. Inside this contour 
$n_B/n_0 > 0.1$ at $t=$ 2, 4, 8 fm/c and $n_B/n_0 > 0.01$ $t=$ 12 fm/c. 
%where $n_B$ is the proper baryon density and $n_0=0.15$ 1/fm$^3$ is 
%the normal nuclear density. 
{
The inner bold dashed contour indicates the freeze-out border. Inside this contour 
the matter still hydrodynamically evolves, while outside -- it is frozen out.
At $t=$ 2 and 4 fm/c there is no frozen-out matter, while at 
$t=$ 12 fm/c all the matter is frozen out.
}
}
\label{fig1}
\end{figure*}
\begin{figure*}[!tbh]
%\vspace*{-14mm}
\includegraphics[width=18.cm]{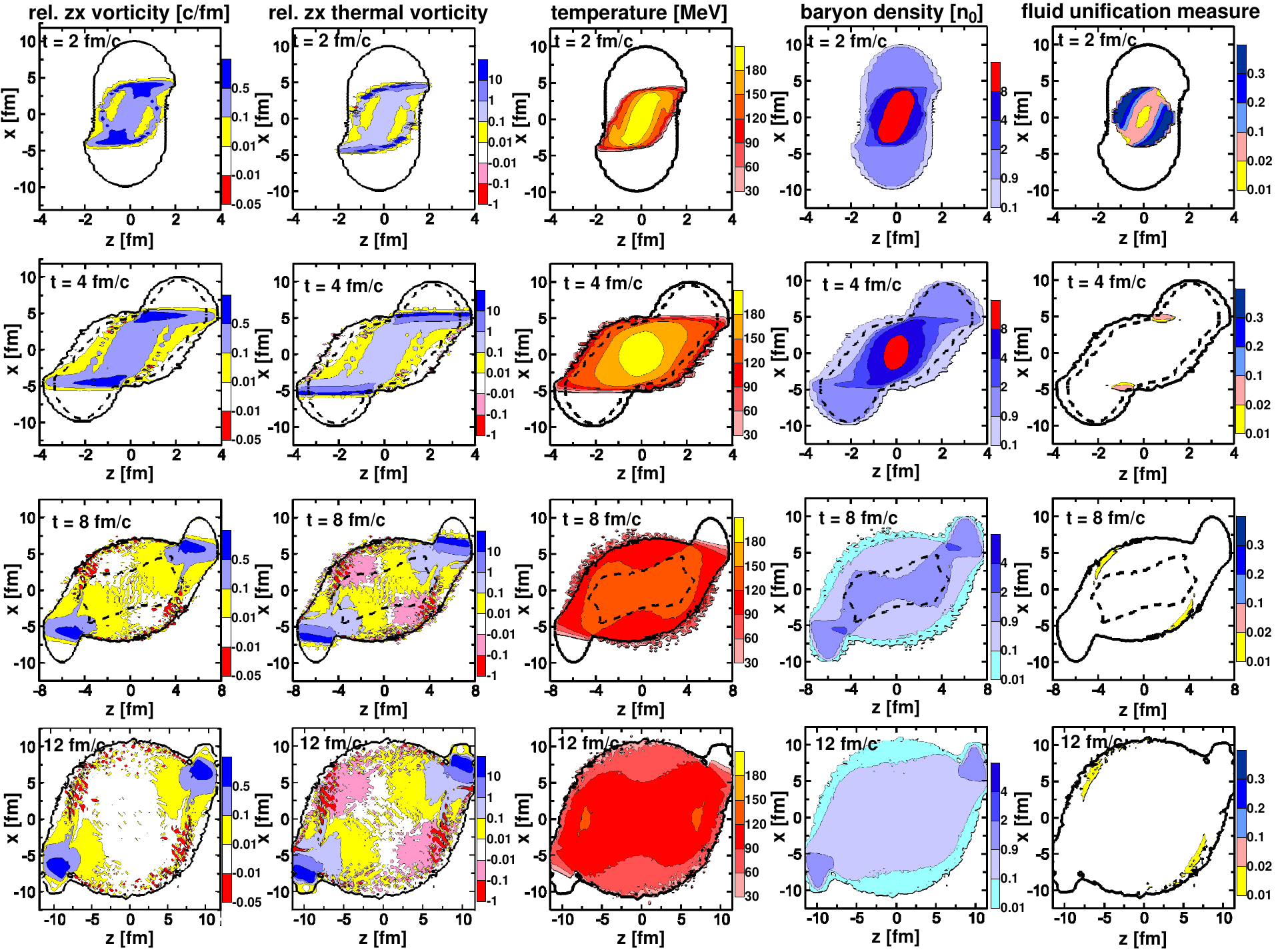}
 \caption{(Color online)
The same as in Fig. \ref{fig1} but at $\sqrt{s_{NN}}=$ 7.7 GeV.
{
At $t=$ 2 fm/c there is no frozen-out matter, while at 
$t=$ 8 fm/c all the matter is frozen out.
}
}
\label{fig2}
\end{figure*}

In Fig. \ref{fig0} the time evolution of the total angular momentum, 
the angular momentum of the baryon-rich fluids in their overlap 
region and the angular momentum of the fireball fluid 
in the semi-central ($b=$ 6 fm) Au+Au collision at $\sqrt{s_{NN}}=$ 7.7 GeV are presented. 
%Calculations are done with the crossover EoS. 
The total angular momentum (that includes a contribution of spectators) is a conserved quantity. 
Therefore, its constancy demonstrates the accuracy of the numeric scheme: $J_{\rm{total}}$
is conserved with the accuracy of 1.5\%. The overlap region rises in the course of 
interpenetration of nuclei and then, at the expansion stage this region includes 
more and more former spectators. Thus, 
the angular momentum of the baryon-rich fluids in their overlap 
almost completely involves the total angular momentum of the system at the final stage of the collision. 
The angular momentum of the newly produced f-fluid is almost 
two orders of magnitude lower than that of the overlapped baryon-rich fluids 
at the considered collision energy. Moreover, the baryon-rich fluids and 
fireball fluid are located in the same rapidity range at the considered collision energy. 
These are additional arguments to neglect the contribution of the fireball fluid 
vorticity in the consideration below.

\section{Results}
\label{Results}

The 3FD simulations of Au+Au collisions were performed 
%at energies $\sqrt{s_{NN}}\le$ 4.9 and 7.7 GeV were performed 
without freeze-out. 
The freeze-out in the 3FD model removes the frozen out matter from the hydrodynamical 
evolution \cite{Russkikh:2006aa}. Therefore, in order to keep all the matter in the 
consideration the freeze-out was turned off.

In order to suppress contributions of almost empty regions, 
where the matter is relatively thin, we consider a  
proper-energy-density-weighted relativistic kinematic  vorticity in the reaction ($xz$) plane, 
i.e. at $y=0$ 
   \begin{eqnarray}
   \label{en.av.rel.B-vort}
   \Omega_{\mu\nu} (x,0,z,t) = \omega_{\mu\nu}(x,0,z,t)  \varepsilon_B (x,0,z,t)
   / \langle \varepsilon_B (0,t) \rangle, 
   \end{eqnarray}
similar to that in Refs. \cite{Csernai:2013bqa,Csernai:2014ywa}. 
Here 
   \begin{eqnarray}
   \label{B-en.av.}
\langle \varepsilon_B (y,t) \rangle = 
\int dx \; dz \; \varepsilon_B (x,y,z,t) 
\Big/
\int_{\varepsilon_B (x,y,z,t)>0} dx \; dz
\cr
   \end{eqnarray}
is  the  energy density of net-baryon-rich fluids, 
$\varepsilon_B=\varepsilon_p + \varepsilon_t$,  
averaged over an $xz$ plane. Similarly to $\Omega_{\mu\nu}$
we define a proper-energy-density-weighted  thermal vorticity 
in the reaction plane, though keep the same notation ($\varpi_{\mu\nu}$) for it. 
 
In Figs. \ref{fig1} and \ref{fig2}, the proper-energy-density weighted 
relativistic kinematic  $zx$ vorticity [cf. Eqs. (\ref{en.av.rel.B-vort}) and (\ref{B-en.av.})] 
and  the thermal $zx$ vorticity, as well as the 
{
temperature and the proper baryon density, Eqs. (\ref{Tm-dissipative}) and (\ref{nb-prop}), 
respectively, 
of the baryon-rich subsystem in the reaction plain ($xz$) at various time instants 
in semi-central ($b=$ 6 fm) Au+Au collisions at $\sqrt{s_{NN}}=$ 4.9 and 7.7 GeV are presented. 
%Calculations are done with the crossover EoS.  
The figures also present 
the fluid unification measure [cf. Eq. (\ref{unification})].  
As it has been already mentioned, the baryon-rich fluids are  
mutually stopped and unified at $t\gsim 4$ fm/c for 4.9 GeV and $t\gsim 2$ fm/c for 7.7 GeV.  
In particular, this means that the temperature and respectively the thermal vorticity 
are poorly defined at earlier time instants.  
When the freeze-out stars (the inner bold dashed contour in Figs. \ref{fig1} and \ref{fig2})
the baryon-rich system has been already completely equilibrated. 
To the last displayed time instant ($t=$ 12 fm/c) the freeze-out has been already completed. 
}

Contrary to Refs. \cite{Csernai:2014ywa,Teryaev:2015gxa}, where results averaged over 
all slices with different $y$ coordinate were presented, we demonstrate plots of 
$\Omega_{\mu\nu}$ and $\varpi_{\mu\nu}$ for the single slice $y=0$, i.e. the true 
reaction plane. It allows us to reveal certain qualitative features of the vorticity 
field.

%
%\begin{figure}[p]
\begin{figure*}[bht]
%\vspace*{-14mm}
\includegraphics[width=7.5cm]{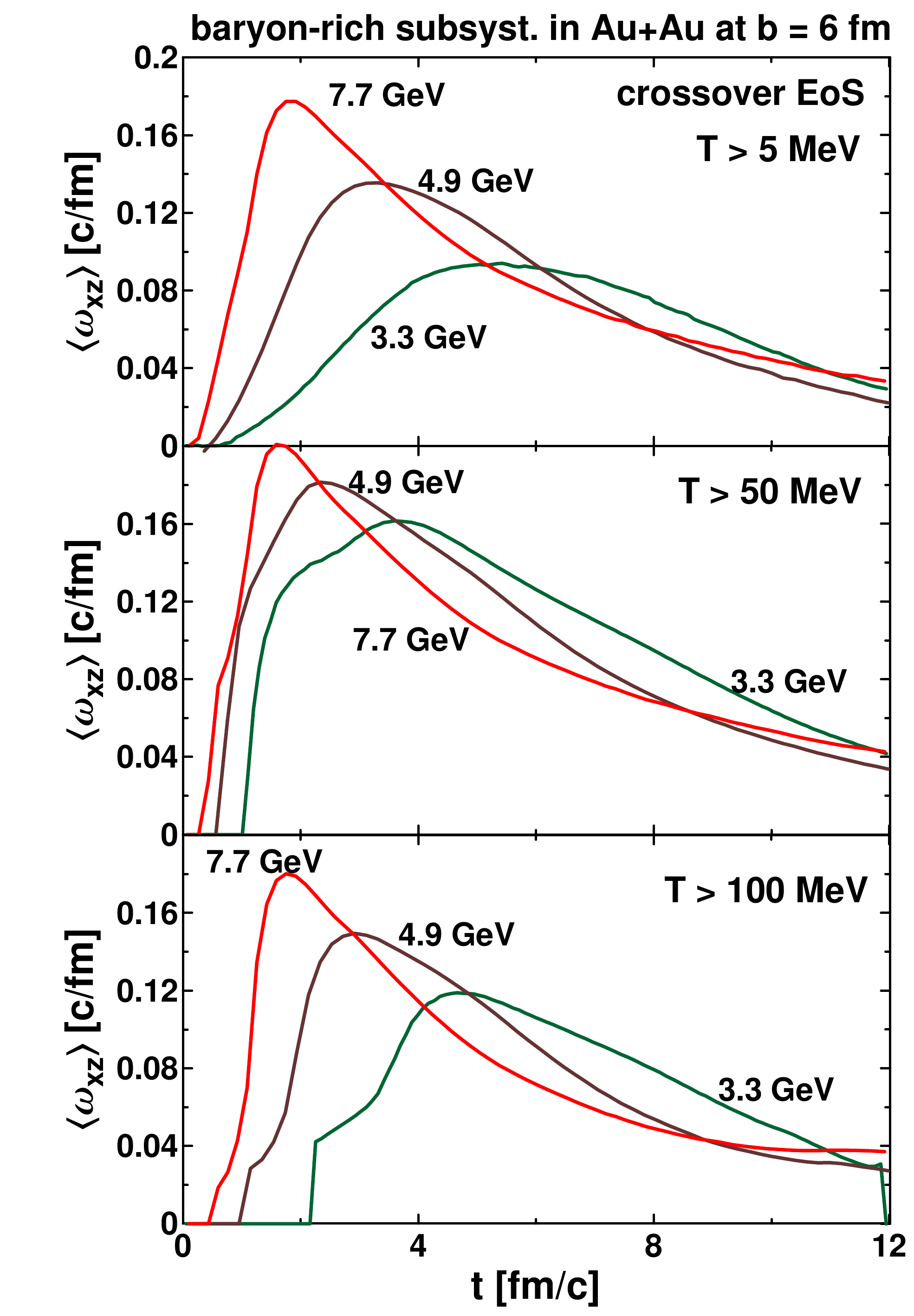}
\includegraphics[width=7.5cm]{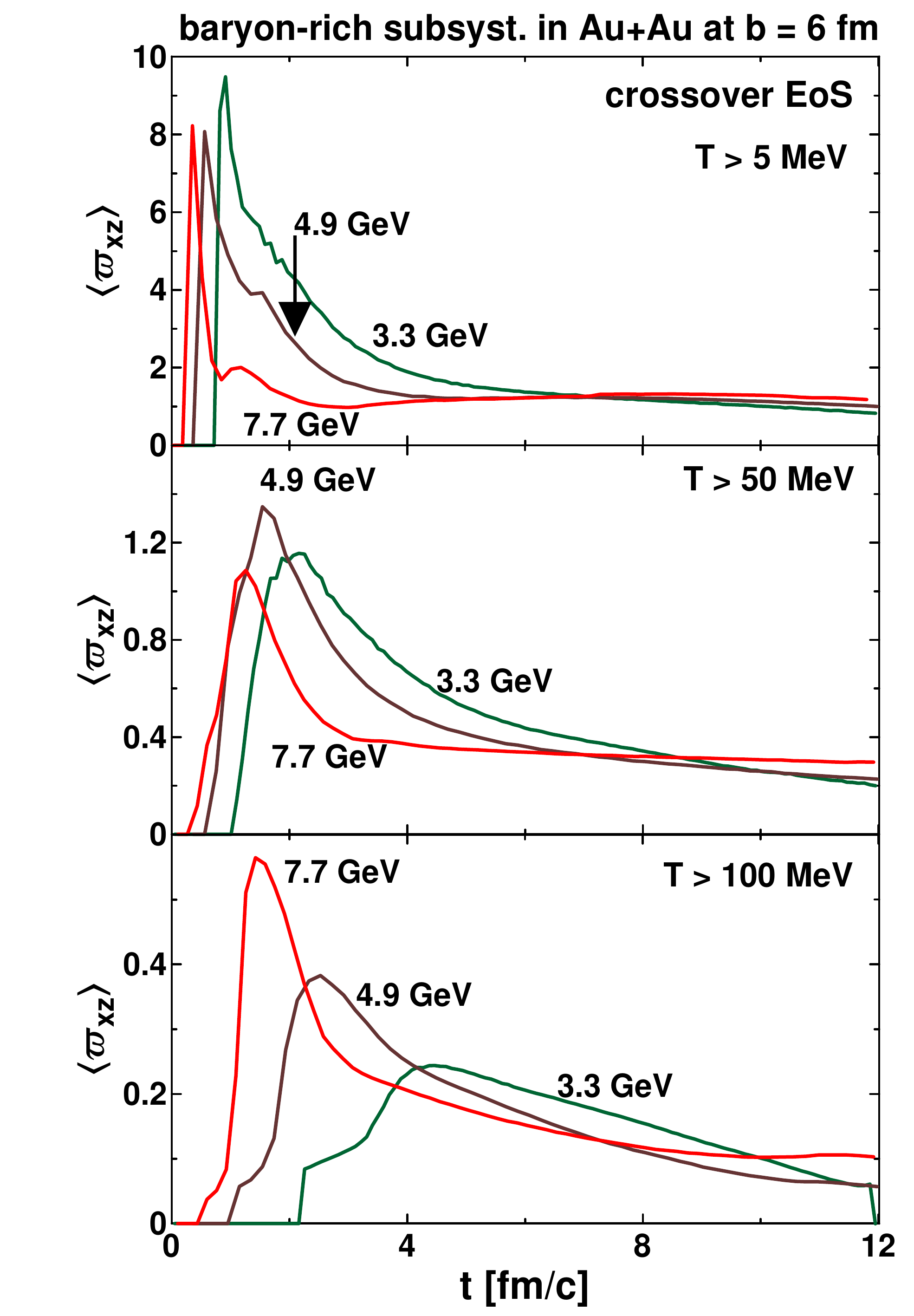}
 \caption{(Color online)
Time evolution of 
relativistic kinematic $zx$ vorticity (left column of panels) and thermal $zx$ vorticity
(right column of panels) of the baryon-rich subsystem 
in the semi-central ($b=$ 6 fm) Au+Au collisions at $\sqrt{s_{NN}}=$ 3.3, 4.9 and 7.7 GeV. 
The vorticities are
averaged with the weight of the proper energy density over 
different regions with temperatures $T>$ 5, 50, and 100 MeV. 
%Calculations are done with the crossover EoS.
}
\label{fig4}
\end{figure*}
As seen, the relativistic kinematic vorticity  and thermal vorticity
primarily start at the border between the participant and spectator matter. 
Later on they partially spread to the participant and spectator bulk though 
remain concentrated near the border.  
In the conventional hydrodynamics this 
extension into the bulk of the system is an effect of the shear viscosity. In the 3FD dynamics it 
is driven by the 3FD dissipation which imitates the effect of the shear viscosity \cite{Ivanov:2016vkw}.
The spread into the bulk is more spectacular at lower collision energy ($\sqrt{s_{NN}}=$ 4.9 GeV) 
because of the higher effective shear viscosity than that at higher energies \cite{Ivanov:2016vkw}. 
At the same time, the vorticity in the participant bulk gradually dissolves in the course of time 
and practically disappears in the center of the colliding system to the end of the collision.   

This observation has consequences for the  polarization of secondary produced particles. 
These particles are abundantly produced in the most dense and hot regions of the system, 
i.e. in the center of the colliding system. However, the vorticity is small there. 
The polarized particles dominantly originate from peripheral regions with high vorticity and 
quite moderate temperature, see right panels in Figs. \ref{fig1} and \ref{fig2}. 
Therefore, we should not expect a large overall polarization of $\Lambda$ hyperons in spite 
of high peak values of the vorticity. At the same time, the relative polarization of $\Lambda$ hyperons
should be higher in the fragmentation regions, i.e. the kinematical region of the participant-spectator border,   
than that in the midrapidity region. 
In the vorticity plots presented in Refs. 
\cite{Csernai:2014ywa,Teryaev:2015gxa} the vorticity occupy the bulk of the participants. 
This happens because of because of the averaging over 
all slices with different $y$ coordinate applied there. This averaging smears out 
the vorticity peaks at the border.

{
As one can see, the peak values of the thermal vorticity reach extremely high values. 
}
This is because of strong gradients of the temperature 
at the border between the participant and spectator matter. These gradients enhance the 
thermal vorticity. The peak values can not be directly compared with those presented in 
Refs. \cite{Csernai:2014ywa,Teryaev:2015gxa} because of the additional averaging over 
all slices with different $y$ coordinate applied there. 
In order to perform a more informative comparison we calculated
relativistic  (kinematic and thermal) $zx$ vorticity  of the baryon-rich subsystem 
in the semi-central ($b=$ 6 fm) Au+Au collision at $\sqrt{s_{NN}}=$ 3.3, 4.9 and 7.7 GeV 
averaged with the weight of the proper energy density over the whole system, 
see Fig. \ref{fig4}. %and \ref{fig5}. 
Keeping in mind that the $\Lambda$ hyperons are abundantly produced 
from the hottest regions of the system, we applied certain constraints on this averaging. 
We considered three regions of the averaging: (i) a region with temperatures $T>5$ MeV that in fact
includes all the participant region, and two regions with more stringent constraints, i.e.  
(ii) $T>50$ MeV and (iii) $T>100$ MeV. These biased averaged quantities can be   
expressed as follows 
   \begin{eqnarray}
   \label{en.av.rel.B-vort-T}
   \langle \omega_{\mu\nu} (t) \rangle_{T>T_0} &=& \int_{T>T_0} dV \;\omega_{\mu\nu}(x,y,z,t)\;\varepsilon_B (x,y,z,t)
   \cr
 &\Big/& \int_{T>T_0} dV \; \varepsilon_B (x,y,z,t)
   \end{eqnarray}
   \begin{eqnarray}
   \label{en.av.therm.B-vort-T}
   \langle \varpi_{\mu\nu} (t) \rangle_{T>T_0} &=& \int_{T>T_0} dV \;\varpi_{\mu\nu}(x,y,z,t)\;\varepsilon_B (x,y,z,t)
   \cr
 &\Big/& \int_{T>T_0} dV \; \varepsilon_B (x,y,z,t)
   \end{eqnarray}
where $T_0$ is the temperature constraint.

Time evolution of the biased 
relativistic kinematic $zx$ vorticity and thermal $zx$ vorticity of the baryon-rich subsystem 
averaged with the weight of the proper energy density over the whole system 
%in the semi-central ($b=$ 6 fm) Au+Au collisions at $\sqrt{s_{NN}}=$ 3.3, 4.9 and 7.7 GeV 
is presented in Fig. \ref{fig4}. % and \ref{fig5}.
As seen, the kinematic vorticity weakly depends on the temperature constraint. 
At the initial (compression) stage of the collision the kinematic vorticity differs at 
different collision energies. However, at the expansion stage, i.e. after the maximum of 
$\langle \omega_{\mu\nu} (t) \rangle$, the kinematic vorticity becomes less sensitive to 
the collision energy. Moreover, the values of the kinematic vorticity are almost independent 
of the collision energy at the final (``freeze-out'') stage, though this concerns quite a narrow 
range of collision energies. These final-stage values are compatible with those at 
$\sqrt{s_{NN}}=$ 5 GeV obtained in Ref.
\cite{Teryaev:2015gxa} within the hadron-string dynamics model \cite{Cassing99}. 
As compared with the results of Ref. \cite{Csernai:2014ywa} within the relativistic 
PICR hydro approach \cite{Csernai:2013bqa}
at $\sqrt{s_{NN}}=$ 8 GeV, the whole expansion PICR stage is very (quantitatively and qualitatively) 
similar to that in our simulations at $\sqrt{s_{NN}}=$ 7.7 GeV.

\begin{figure}[hbt]
\includegraphics[width=7.5cm]{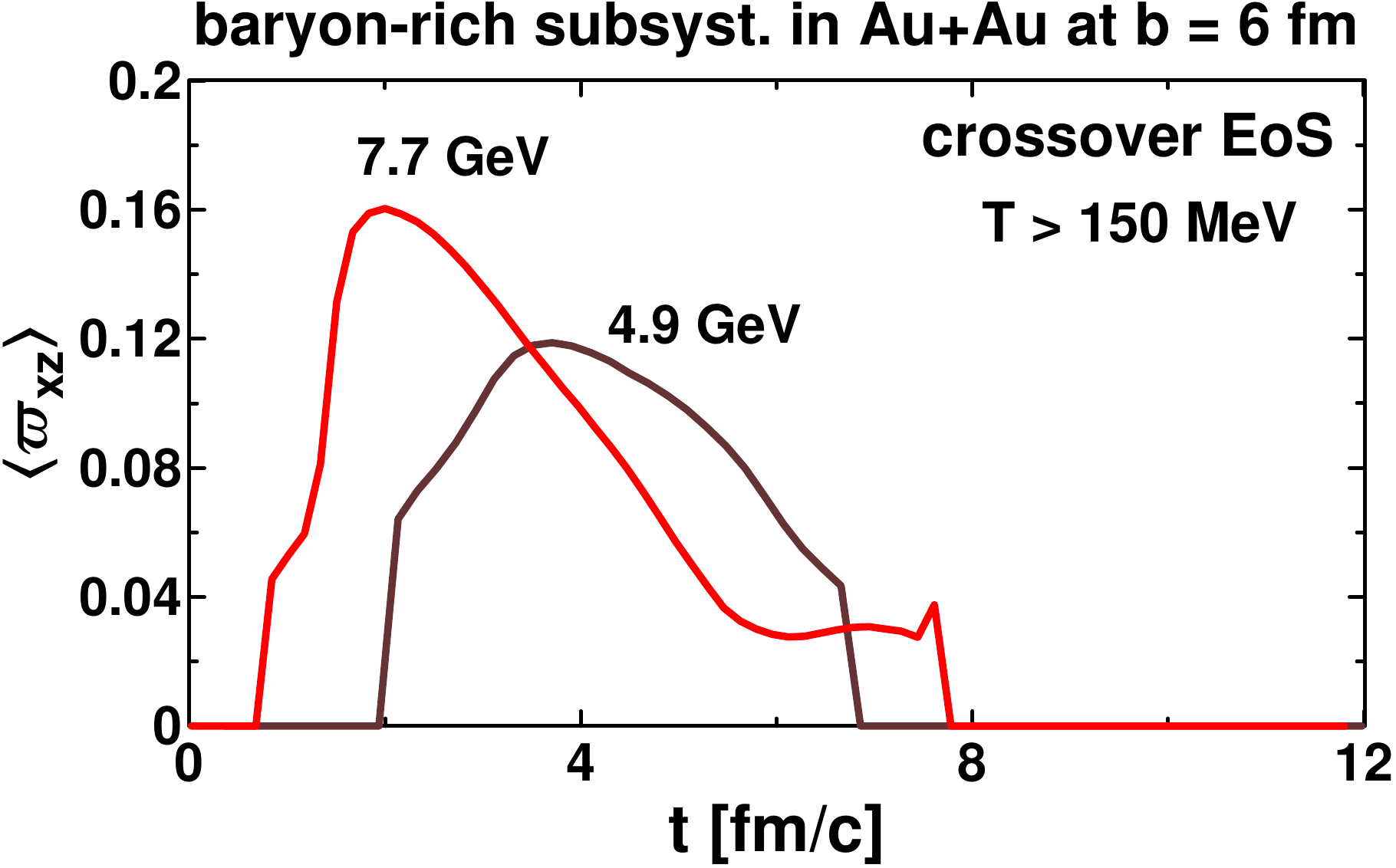}
 \caption{(Color online)
The same as in Fig. \ref{fig4} but for the relativistic 
thermal $zx$ vorticity with the constraint $T>$ 150 MeV.  
 }
\label{fig6}
\end{figure}

{
It is worthwhile to mention that the averaged vorticity displayed in Fig. \ref{fig4} 
does not coincide with that of the frozen-out system. 
The freeze-out in the 3FD model is a continuous in time process \cite{3FD,Russkikh:2006aa}, 
as it is illustrated in Figs. \ref{fig1} and \ref{fig2}. 
When the the freeze-out happens the frozen-out matter stops to hydrodynamically evolve. 
In particular, the achieved vorticity also turns out to be frozen out. 
In the calculation presented in Fig. \ref{fig4} all the matter hydrodynamically evolves 
without exemptions till the very late time. Therefore, the late-stage values presented in Fig. \ref{fig4} 
can be considered only as an estimate of the vorticity averaged over the frozen-out system. 
}

{
Enormously high peak values of the thermal vorticity, 
see the right column of panels in Fig. \ref{fig4}, 
are irrelevant because of the above mentioned 
poor definition of this vorticity at the early stages of the collision, 
i.e. at $t\lsim 4$ fm/c for 4.9 GeV and $t\lsim 2$ fm/c for 7.7 GeV.  
}
The relativistic thermal $zx$ vorticity averaged with the weight of the proper energy density 
over the whole system 
exhibits features similar to those observed in the relativistic kinematic $zx$ vorticity 
except that  $\langle \varpi_{\mu\nu} (t) \rangle$ strongly depends on the temperature constraint. 
This a consequence of the cutoff of near-spectator regions with high temperature gradients  
at high-$T_0$ constraint, see Eq. (\ref{en.av.therm.B-vort-T}). Nevertheless, even at $T_0=100$ MeV 
cutoff the $\langle \varpi_{\mu\nu} (t) \rangle$ values at $\sqrt{s_{NN}}=$ 7.7 GeV essentially exceed
those reported in Ref. \cite{Csernai:2014ywa} for $\sqrt{s_{NN}}=$ 8 GeV. 
This happens because the near-spectator regions still contribute even at $T_0=100$ MeV cutoff. 
Only at $T_0=150$ MeV our $\langle \varpi_{\mu\nu} (t) \rangle$ values become comparable 
with those of Ref. \cite{Csernai:2014ywa}, see Fig. \ref{fig6}.    
In view of that the kinematic $zx$ vorticity is 
well comparable within the present 3FD and PICR-hydro \cite{Csernai:2013bqa} approaches, 
we can conclude that the temperature gradients in the periphery of the participant zone 
are much stronger in the 3FD model.

%min. total energy of the baryon-rich fluids 1.399E+03 GeV [-$2Am_N$ = 370 GeV] 

%max. total energy of the baryon-free fluid 9.994E+01 GeV 

\section{Summary}
\label{Summary}

Within the 3FD model (crossover scenario) we have studied vorticity evolution 
in heavy-ion collisions at NICA energies. We considered two definitions of the 
vorticity---relativistic kinematic and thermal vorticities---that are relevant in different 
aspects of the rotation effects.

It is found that 
the vorticity  mainly takes place at the border between participant and spectator matter. 
This effect was noticed in the analysis of the kinematic vorticity field 
\cite{Baznat:2015eca,Baznat:2013zx} in the framework of the kinetic Quark-Gluon String Model. 
The authors of  Refs. \cite{Baznat:2015eca,Baznat:2013zx} observed 
that the vorticity field is predominantly localized in a relatively thin layer 
on the boundary between participants and spectators and that it
forms a specific toroidal structure---a so called femto-vortex sheet.  
As we found, this effect is essentially enhanced for the case of 
the thermal vorticity because of strong temperature 
gradients at the participant-spectator border. 
As the thermal vorticity is directly related to the $\Lambda$-hyperon polarization, 
this implies that 
the $\Lambda$-hyperon polarization should be stronger at peripheral rapidities,  
corresponding to the participant-spectator border, 
than that in the midrapidity region.

At the expansion stage of the collision the vorticity is only weakly dependent 
on the collision energy, though the considered NICA energy range is quite narrow. 
The order of magnitude of the mean weighted kinematic vorticity
agree with that estimated in Ref. \cite{Teryaev:2015gxa} (for $\sqrt{s_{NN}}=$ 4.9 GeV) and  
in Ref. \cite{Csernai:2014ywa} (for $\sqrt{s_{NN}}=$ 8 GeV). 
At the same time, obtained values of the mean weighted thermal vorticity at $\sqrt{s_{NN}}=$ 7.7 GeV, 
which is directly related to the $\Lambda$-hyperon polarization, are an order of magnitude higher than 
those reported in Ref. \cite{Csernai:2014ywa} for $\sqrt{s_{NN}}=$ 8 GeV. 
Additional constraint to high-temperature 
($T>$ 150 GeV) participant region, over which the mean values are calculated, 
reduces the mean values of the thermal vorticity by an order of magnitude 
and makes them comparable with those found in Ref. \cite{Csernai:2014ywa}. 
Only this strong constraint ($T>$ 150 GeV) excludes the effect of strong temperature 
gradients at the participant-spectator border.

\vspace*{3mm} {\bf Acknowledgments} \vspace*{2mm}

Fruitful discussions with D.N. Voskresensky 
are gratefully acknowledged. 
We are also grateful to Oleg Teryaev for valuable comments on the manuscript of the paper. 
The calculations were performed at the computer cluster of GSI (Darmstadt). 
Y.B.I. was supported by the Russian Science
Foundation, Grant No. 17-12-01427.
A.A.S. was partially supported by the Academic Excellence Project of 
the NRNU MEPhI under contract 
%with the Ministry of Education and Science of the Russian Federation 
No. 02.A03.21.0005. %, 27.08.2013. 

\end{document}